\documentclass{emulateapj}
\usepackage{apjfonts}

\newcommand\chandra{{\it Chandra}}
\newcommand\eg{{e.g.}}
\newcommand\etal{{et al.}}
\newcommand\cf{{cf.}}
\newcommand\msun{{\rm\ M_\odot}}
\newcommand\ergps{{\rm\ erg\ s^{-1}}}

\begin{document}

\title{The powerful outburst in Hercules A}

\author{P. E. J. Nulsen\altaffilmark{1,2}, 
  D. C. Hambrick\altaffilmark{1,3},  B. R. McNamara\altaffilmark{4},
  D. Rafferty\altaffilmark{4}, L. Birzan\altaffilmark{4},
  M. W. Wise\altaffilmark{5},   L. P. David\altaffilmark{1}}
\altaffiltext{1}{Harvard-Smithsonian Center for Astrophysics, 60
  Garden Street, Cambridge, MA 02138; pnulsen@cfa.harvard.edu}
\altaffiltext{2}{On leave from the University of Wollongong}
\altaffiltext{3}{Harvey Mudd College, 301 E. Foothill Blvd.,
  Claremont, CA 91711}
\altaffiltext{5}{Astrophysical Institute, and Department of Physics
  and Astronomy, Ohio University, Clippinger Laboratories, Athens, OH
  45701}
\altaffiltext{5}{Center for Space Research, Building NE80-6015,
  Massachusetts Institute of Technology, Cambridge, MA 02139}

\begin{abstract}

The radio source Hercules A resides at the center of a cooling flow
cluster of galaxies at redshift $z = 0.154$.  A \chandra\ X-ray image
reveals a shock front in the intracluster medium (ICM) surrounding the
radio source, about 160 kpc from the active galactic nucleus (AGN)
that hosts it.  The shock has 
a Mach number of 1.65, making it the strongest of the cluster-scale
shocks driven by an AGN outburst found so far.  The age of the
outburst $\simeq 5.9\times10^7$ y, its energy $\sim 3\times10^{61}$
erg and its mean power $\sim 1.6\times10^{46} \ergps$.  As for the
other large AGN outbursts in cooling flow clusters, this outburst
overwhelms radiative losses from the ICM of the Hercules A cluster by
a factor of $\sim100$.  It adds to the case that AGN outbursts are a
significant source of preheating for the ICM.  Unless the mechanical
efficiency of the AGN in Hercules A exceeds $10\%$, the central black
hole must have grown by more than $1.7\times 10^8 \msun$ to power this
one outburst.

\end{abstract}

\keywords{cooling flows -- galaxies: clusters:
  individual (\objectname{Hercules A}) -- intergalactic medium -- X-rays:
  galaxies: clusters}

\section{Introduction}

It is not yet clear which heating mechanism
\citep[\eg][]{nm01,mbl04} is chiefly responsible for preventing gas
from cooling in cluster cooling flows
\citep{pkp03}, but the most promising is heating by a central AGN 
\citep{tb93,td97}.  Heating and cooling rates are linked if the AGN is
fed by cooled or cooling gas.
Such feedback could maintain otherwise unstable
cool cores, explaining the prevalence of cooling flows
\citep{f94}.  AGN powered radio lobe cavities
\citep[\eg][]{cph94,mwn00,fse00} heat the ICM \citep{csf02},
but not enough to make up for radiative losses
\citep{brm04}.  Weak ``cocoon'' shocks, long expected in models of
jet-fed radio lobes \citep[\eg][]{s74,hrb98}, have been found in a
number of systems \citep{fsa03,fnh04,mnw04,nmw04}.  They represent
additional heating due to AGN outbursts, and provide a new tool for
determining ages and energies of AGN outbursts.

This letter reports the discovery of a shock front generated by the
AGN outburst that powers Hercules A.  One of the
brightest radio sources in the sky \citep{df84,gl03}, Hercules A
resides at the center of a cluster of galaxies with X-ray
luminosity $\simeq 5\times10^{44}\ergps$, at redshift $z=0.154$
\citep{skb99,gl04}.  Despite its high radio luminosity, Hercules A
lacks bright radio hotspots and so belongs to Fanaroff-Riley class I,
but with an unusual, jet-dominated, radio morphology
\citep{df84,gl03}.  Using Einstein spectra, \citet{wjf97} found a
formal cooling rate of zero for the Hercules A Cluster, but the high
central density we find ($n_{\rm e} \gtrsim 0.1\rm\ cm^{-3}$) gives it
a central cooling time typical of a cooling flow cluster.

Section \ref{sec:obs} gives details of the observations and data
reduction, and section \ref{sec:struct} discusses the main
features of the \chandra\ image of Hercules A.  Properties of the
shock are discussed in section \ref{sec:shock} and its
implications in section \ref{sec:disc}.  Flat
$\Lambda$CDM, with $H_0 = 70\rm\ km\ s^{-1}\ Mpc^{-1}$ and
$\Omega_{\rm m} = 0.3$, is assumed throughout, giving a scale of
$2.67\rm\ kpc\ arcsec^{-1}$ for Hercules A.

\section{Observations and data reduction} \label{sec:obs}

Hercules A was observed with \chandra\ for 14.8 ksec on 25 Jul 2001,
in VFAINT mode with ACIS-S at the aim point (OBSID 1625).  For the
analysis here, the event list was reprocessed using recent
calibrations.  It was screened to remove
 ASCA grades 1, 5 and 7, and
bad pixels.  Periods of high particle background were removed
following the method of
Markevitch\footnote{http://hea-www.harvard.edu/~maxim/axaf/acisbg},
leaving 12.4 ksec of good exposure time.  After cleaning, the mean
count rate in ACIS S1 was $0.136\rm\ ct\ s^{-1}$, $\sim 3.5\sigma$
(9\%) higher than
expected\footnote{http://hea-www.harvard.edu/~maxim/axaf/acisbg/data/README},
suggesting some residual contamination due to particle
background.  Data were processed to correct for time dependence of the
ACIS
gain\footnote{http://hea-www.harvard.edu/{${\sim}$}alexey/acis/tgain/}
and filtered according to the prescription of
Vikhlinin\footnote{http://cxc.harvard.edu/cal/Acis/Cal\_prods/vfbkgrnd/}
to reduce particle background.  Background event files were created by
processing standard ACIS background files in the same manner as
the data.  Point sources were identified manually for removal from
spectra and surface brightness profiles.
ARF's and RMF's for extended regions are weighted by number of events.
ARF's are corrected to allow for the reduction in low energy response
due to contaminant on the ACIS filters.

\section{The X-ray image of Hercules A} \label{sec:struct}

\begin{figure}
\centerline{\includegraphics[width=\linewidth]{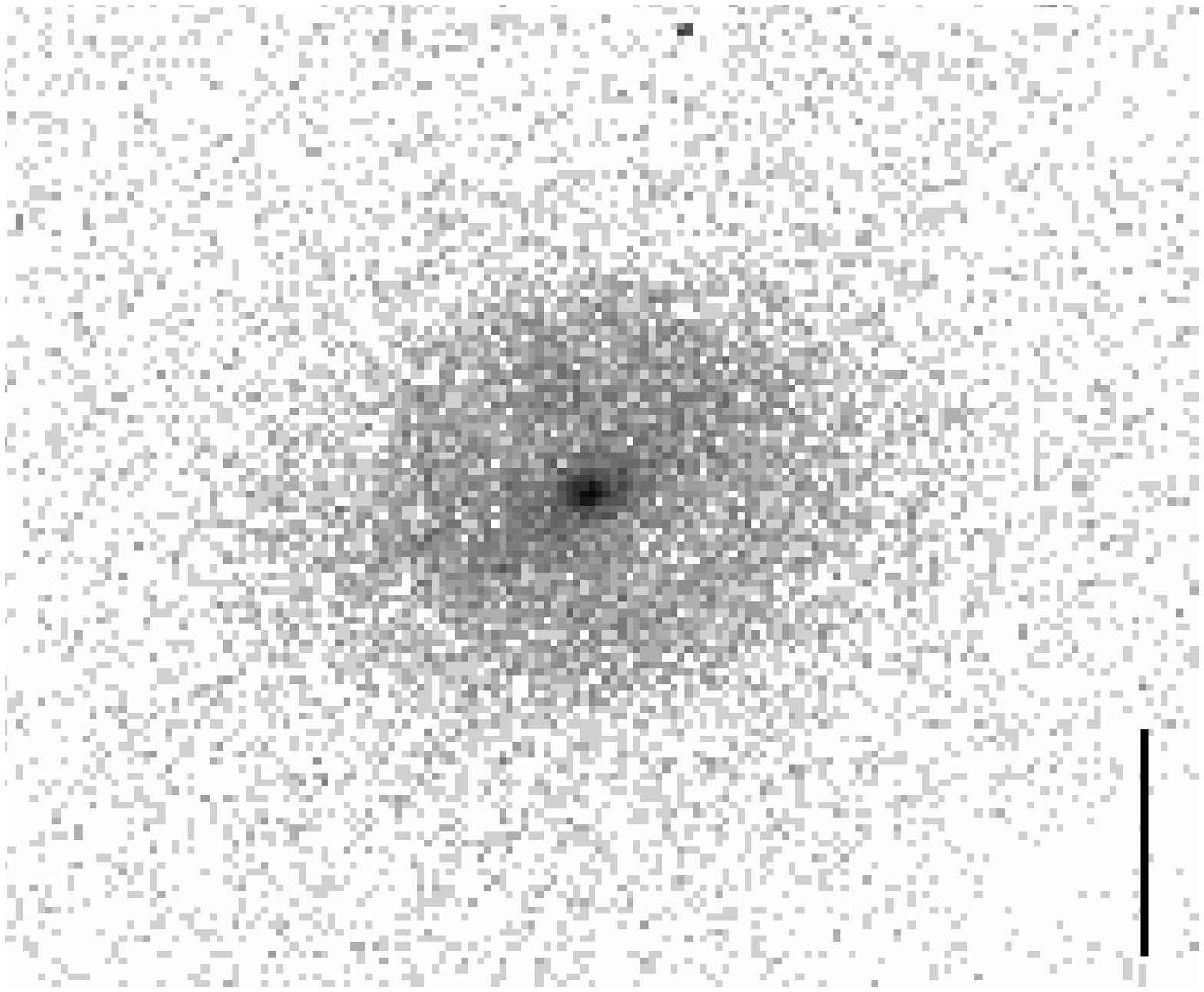}}
\centerline{\includegraphics[width=\linewidth]{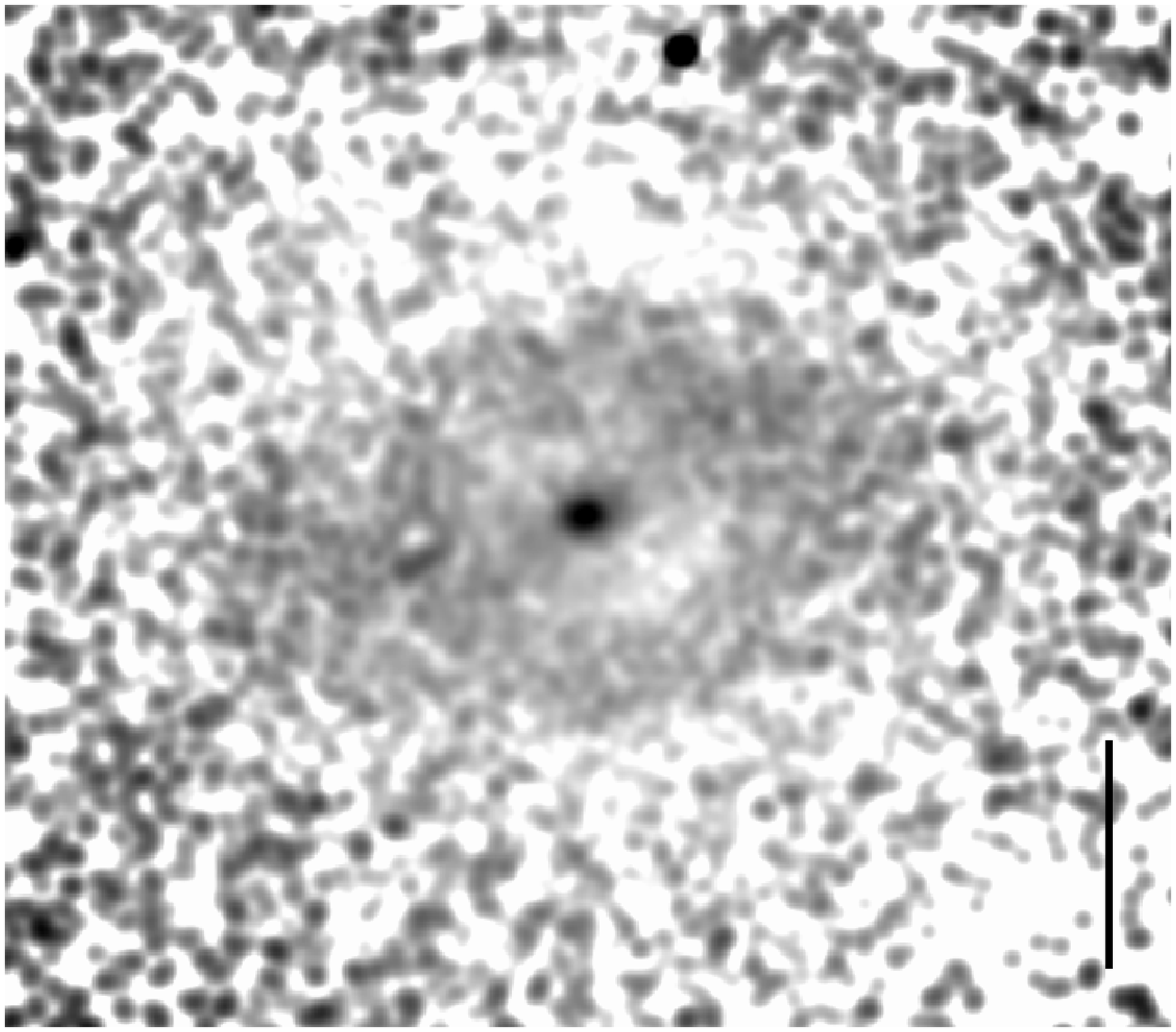}}
\caption{{\it Upper panel:} 0.3 -- 7.5 keV \chandra\ image of Hercules
  A made from the distributed evt2 file binned by a factor of 4.  {\it
  Lower panel:} 0.3 -- 7.5 keV image of Hercules A made from the
  cleaned, reprocessed data, smoothed with a $2''$ gaussian and
  divided by a beta model.  The scale bar in each panel is 1 arcmin
  (160 kpc) in length.  The bright central region $\sim1$ arcmin in
  radius is surrounded by the shock front.  The southwest cavity is
  $\sim 0'\!.5$ from the central peak.}
\label{fig:rawsm}
\end{figure}

The image in the upper panel of Fig.~\ref{fig:rawsm} shows the filtered
and calibrated events from the raw \chandra\ data (evt2 file) for
Hercules A, binned by
a factor of 4.  The lower panel shows a 0.3 - 7.5 keV image
made after the cleaning and reprocessing described
above.  The image has been smoothed with a $2''$ gaussian and
divided by a beta model, with $42''$ core radius and a beta
of 0.6, centered on the X-ray peak \citep[core radius
from][but smaller beta]{gl04}.
Division by the beta model reduces the radial variation of surface
brightness, making it easier to discern substructure over a
substantial range of radius.  Each image has a $1'$ scale bar.

\begin{figure}
\includegraphics[width=\linewidth]{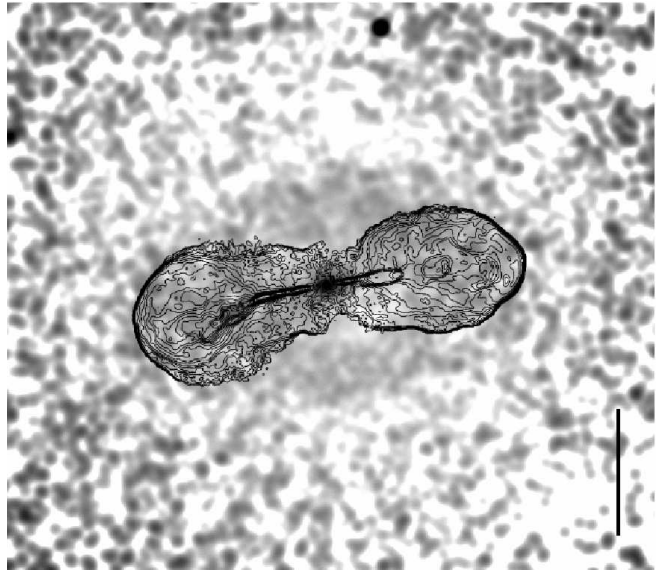}
\caption{X-ray and radio images of Hercules A.  The \chandra\ image of
Fig.~\ref{fig:rawsm} overlaid with 1.4 GHz radio contours from
\citet{gl03}.}
\label{fig:radio}
\end{figure}

Although the central peak of the X-ray image is prominent, it is well
resolved by \chandra\ and there is no sign of a point-like AGN
\citep[\cf][]{tfm01}.
A striking feature of the X-ray image is the bright region,
roughly $1'$ in radius, that stands out in the upper panel of
Fig.~\ref{fig:rawsm}.  This has a similar size to the radio emission
and extends to the east and west around the radio
lobes (Fig.~\ref{fig:rawsm} lower, Fig.~\ref{fig:radio}).  Its shape
and association with the radio 
source suggest it is the shocked cocoon of the
expanding radio lobes \citep{s74,hrb98}.  The break in surface
brightness that bounds this region is shown to be consistent with a
shock front below.

There is a $\sim 7 \sigma$ deficit of X-ray emission in the region
$\sim0'\!.5$ to the  
southwest of the bright center in Fig.~\ref{fig:rawsm}, $\sim15''$ (40
kpc) in radius.  There is a weaker, $\sim 3\sigma$, deficit in the
X-ray emission from
the corresponding region to the northeast, partly
masked by a bright spot of X-ray emission to the north of the center.
These features resemble the cavities associated with many other
cluster radio sources \citep[\eg][]{mwn00,fse00}.  However, they are
not aligned with the axis of the radio jets and do not contain
radio lobes.  They might be ghost cavities \citep[\eg][]{mwn01}, but
if so, it is surprising that they lie within an active radio source.

Lastly, there is a ridge of enhanced X-ray emission crossing the
the bright region, from $\sim30^\circ$ south of east to
$\sim30^\circ$ north of west, roughly at right angles to the axis
defined by the cavities.  This feature also has no obvious association
with the radio source (it forms an angle of $\sim20^\circ$ with the
radio jets).  The excess emission appears to be thermal, due to
relatively cool, dense gas, which cannot be fully supported by
hydrostatic forces.  The gas may be cool filaments, like those seen in
other cluster central radio sources
\citep[\eg][]{fnh04,nmw04}, or it may be a cooler disk that is
partly supported by rotation.

\section{The Shock Front in Hercules A} \label{sec:shock}

\begin{figure}
\centerline{\includegraphics[height=\linewidth,angle=270]{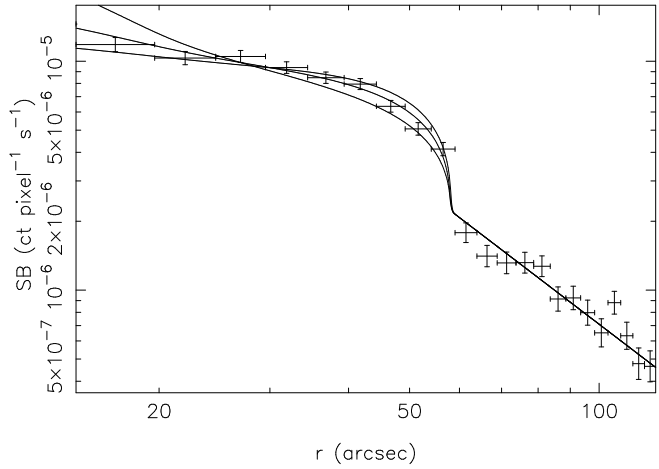}}
\caption{Surface brightness profile of the shock front in Hercules A.
  The 0.6 -- 7.5 keV surface brightness profile is measured in sectors
  from PA $330^\circ$ to $50^\circ$ and $150^\circ$ to $230^\circ$, to
  the north and south of the AGN, at right angles to the jet axis.
  Surface brightness errors are $1\sigma$ statistical errors.  Radial
  error bars show the limits of the bins.  The smooth curves are
  surface brightness profiles for shock models with Mach numbers of
  1.51, 1.65 and 1.79, from bottom to top on the right.  Models are
  scaled to match the observed surface brightness outside the shock.}
  \label{fig:sbns}
\end{figure}

The surface brightness profile of the bright circular region was
measured in two $80^\circ$ sectors, approximately at right angles to
the axis of the radio jet.  This avoids smearing the edge in the
surface brightness profile due to elongation of the bright region to
the east and west.  Fig.~\ref{fig:sbns} shows the radial surface
brightness profile for the ranges of PA $330^\circ$ -- $50^\circ$ and
$150^\circ$ -- $230^\circ$ combined.  Point sources were eliminated,
background subtracted and the resulting profile exposure corrected.
Although the data are quite noisy, there is a clear break in surface
brightness at a radius of $60''$ ($\sim160$ kpc), at the edge of the
bright central region.  Beyond the break, the surface brightness is
well fitted by the power law, $r^{-\alpha}$ with $\alpha =
2.11\pm0.43$ (90\%).  We now consider the interpretation of this front
as a shock.

To determine the strength of the shock we use a spherically symmetric,
hydrodynamic model of a point explosion at the center of an initially
isothermal, hydrostatic atmosphere.  Before passage of the shock, the
gas density is assumed to follow the power law, $\rho(r) \propto
r^{-\eta}$, with $\eta = 1.55$, chosen to make the surface brightness
profile of the undisturbed gas match the observed
profile outside the shock.  The gravitational field ($g \propto 1/r$)
is scaled to make the undisturbed atmosphere hydrostatic.  The surface
brightness profile is determined from the model, assuming that the
temperature of the unshocked gas is 4 keV (see below).  Relative
\chandra\ count rates in the 0.6 -- 7.5 keV band are computed using
detector response files from near to the aim point for
these observations.  The XSPEC $\rm wabs \times mekal$ spectral model
was used, with a foreground column density of $6.4\times10^{20}\rm\
cm^{-2}$, a redshift of 0.154 and abundances of 0.5 times solar,
appropriate for Hercules A (model surface brightness profiles are
insensitive to these and the preshock temperature in the relevant
temperature range).  The model is
self-similar, allowing it to be scaled in radius to match the location
of the shock and in normalization to match observed surface brightness
outside the shock.

In Fig.~\ref{fig:sbns} we show surface brightness profiles for model
shocks with Mach numbers of 1.51, 1.65 and 1.79.
A Mach 1.65 shock gives a reasonable fit to the data.  Apart
from the scaling, model parameters (the initial density power-law,
$\eta$, and preshock temperature) are constrained by observations,
leaving only the Mach number of the shock free in the fit.  The model
has a number of shortcomings \citep[the actual outburst is aspherical,
does not inject energy in a single explosion and the initial gas
density is not a power law,][]{nmw04}, so that it can only be expected
to match the data over a limited range of radius behind the shock.
Nevertheless, the fit provides a stringent test that this feature is
due to a shock propagating into the cluster.

\begin{figure}
\includegraphics[height=\linewidth,angle=270]{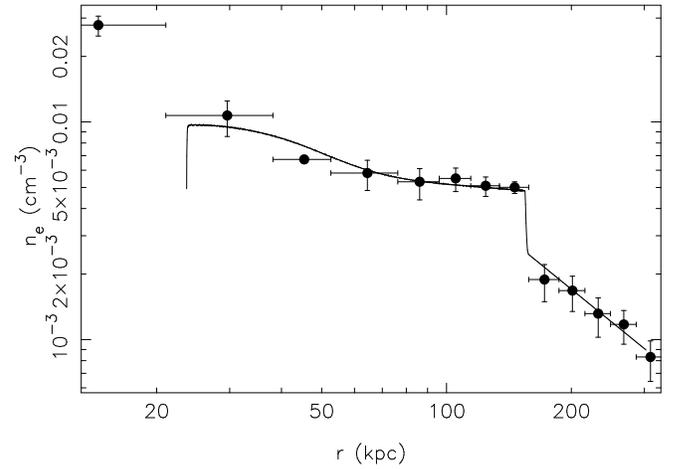}
\caption{Electron density profile of the shock front in Hercules A.
  Deprojected electron density versus radius in the PA ranges
  $330^\circ$ to $50^\circ$ and $150^\circ$ to $230^\circ$.  The shock
  is at 158 kpc.  Density error bars are 90\% confidence ranges.  The
  continuous line shows the density profile for the best fitting
  model.}
  \label{fig:dep} 
\end{figure}

In order to determine physical properties of the outburst from the
model, we must determine the density and temperature of the unshocked
gas.  However, outside the shock from $1'$ to $2'\!.5$, in the sectors
of the surface brightness profile, there are only $\sim1250$ photons
in the 0.6 -- 7.5 keV band.  We have
therefore used a single spectrum extracted from this region to
determine the temperature and normalize the density of the unshocked
gas.  Using an absorbed mekal model, with $N_{\rm H} =
6.4\times10^{20}\rm\ cm^{-2}$, redshift $z=0.154$ and the abundance
set to 0.5, gives a temperature of $kT = 3.9_{-0.6}^{+0.8}$ keV
(90\%).  This is consistent with previous measurements
\citep[\eg][]{gl03}, suggesting that the spectrum is not significantly
affected by the particle background.  Assuming that the gas is
spherically symmetric and its density $\rho(r) \propto r^{-1.55}$ from
the shock to infinity, the normalization of the spectral fit gives an
electron density of $n_{\rm e} = 1.06\pm0.03 \times 10^{-3}\rm\
cm^{-3}$ at a radius of 276 kpc ($1'\!.72$).

Using these parameters, the radius of the shock is 158 kpc, the time
since the outburst is $t_{\rm s} = 5.9\times10^7$ y and the total
energy of the outburst is $E_{\rm s} = 3\times10^{61}$ erg.  This
energy is similar to the lobe enthalpy \citep{gl04}, as expected if
the lobes drive the shock.  The main
source of uncertainty in the age of the outburst ($\sim10\%$) is due
to the uncertainty in the preshock temperature.  The shortcomings of
the model do give rise to systematic uncertainty in the shock energy,
but this is unlikely to be more than a factor $\sim2$
\citep{nmw04}.

In the temperature range 1.6 -- 10 keV, the \chandra\ count rate in
the band 0.6 - 7.5 keV is very insensitive to gas temperature, varying
$\pm 3.3\%$ about its mean over the whole range of
temperature, for a fixed emission measure.  This enables us to
deproject the gas density with reasonable accuracy, despite poor
knowledge of the gas temperature (doubling the abundance to 1.0
would reduce the electron density by $\sim 7\%$).  A deprojection was
done, using the 
method of \citet{nmw04}, with the gas temperature fixed at 4 keV and
other parameters as above.  The resulting electron density profile is
shown in Fig.~\ref{fig:dep}, together with the electron density
profile obtained from the Mach 1.65 shock model.  The results agree
well with the model, clearly showing the density jump at the shock.
The failure of the model for $r \lesssim 25$ kpc is a numerical
artifact, but other shortcomings are expected to make the model
inaccurate at small radii.

In the models, adiabatic expansion limits the size of the region where
the temperature of the shocked gas exceeds that of the unshocked gas.
Nevertheless, the strength of the shock in Hercules A makes it a good
candidate for detecting the temperature rise due to the shock.  For
the Mach 1.65 model, after projection onto the sky, the emission
measure weighted temperature exceeds the preshock temperature by at
least 20\% for 110 kpc $< r <$ 150 kpc.  With other fit parameters as
above, a spectrum extracted from this region ($\sim900$ 0.6 -- 7.5
keV source counts) gives a temperature of $kT = 6.1_{-1.2}^{+2.0}$ keV
(90\%), in reasonable agreement with the model.

\section{Discussion} \label{sec:disc}

The mean power of the outburst in Hercules A, $P_{\rm s} = E_{\rm s} /
t_{\rm s} = 1.6\times10^{46} \ergps$, is two orders of magnitude
larger than the total power radiated from the region where the cooling
time is shorter than $10^{10}$ y.  Hercules A joins a small collection
of cooling flow clusters known to have large-scale shocks driven by an
outburst from an AGN at the cluster center
\citep{fsa03,fnh04,mnw04,nmw04}.  Three of these systems, Hydra A,
MS0735.6+7421 and Hercules A, have outburst energies of
$\gtrsim10^{61}$ erg.  The outburst in Hercules A currently has the
strongest shock and its total energy is the second largest known
\citep[MS0735.6+7421 is twice as energetic,][]{mnw04}.  Along with the
other systems, it has important implications for the energetics of
cooling flows, the preheating of clusters, the interaction of radio
sources with the ICM and the growth of nuclear black holes
\citep{mnw04,nmw04}.

If the outburst is powered by accretion onto a black hole, then the
outburst energy is $E_{\rm s} = \epsilon M_{\rm s} c^2$, where the
mass $M_{\rm s}$ was accreted to fuel the outburst and $\epsilon$ is
the efficiency of jet energy production by the black hole.
Unless $\epsilon > 10\%$, the mass swallowed by the black hole exceeds
$1.7\times10^8 \msun$ to fuel this outburst.  If this mass was
swallowed in a time comparable to the age of the outburst, $t_{\rm s}
\simeq 6\times10^7$ y, the mass increase is hard to reconcile with a
tight correlation between bulge properties and black hole mass
\citep{gbb00}, unless the black hole is very massive indeed.

In our Mach 1.65 model, the shock inverts the entropy profile of gas
inside 58 kpc ($22''$), creating a buoyant bubble at the cluster
center that would then rise.  A large bubble can rise at a significant
fraction of the sound speed \citep{cbk01}, but always more
slowly than the shock front.  Although the shock model is not expected
to match reality closely, the cavities in Hercules A are comparable in
size to the entropy inversion of the model, suggesting that this is
how they were formed.  This would explain the lack of radio
emission from the cavities.

\section{Conclusions} \label{sec:conc}

Analysis of a \chandra\ X-ray image of the Hercules A cluster shows
that it has cavities and a shock front associated with the powerful
radio source.  Unusually, the cavities show no clear connection to
the radio source.  The shock front is elongated in the direction of
the radio lobes and appears to be its cocoon shock.  Fitting a simple
hydrodynamic model to the surface brightness profile gives a Mach
number for the shock of $\simeq 1.65$.  The age of the outburst that
drove the shock is $5.9\times10^7$ y and its total energy is
$3\times10^{61}$ erg.  The deprojected density profile is consistent
with the shock model and, in particular, with the density jump at the
shock.  Within the limits of the spectroscopic data, the temperature
jump is also consistent with the shock model.

The shock outburst is highly significant for the energetics of any
cooling flow in Hercules A and for the cluster as a whole.  The mean
mechanical power of the outburst $\simeq 1.6\times10^{46}\ergps$, well
in the range of quasar luminosities.  The black hole in the AGN that
drove this outburst grew by, at least, $1.7\times10^8 \msun$ during
the outburst.

\acknowledgments

We gratefully acknowledge the assistance of Alexey Vikhlinin and Maxim
Markevitch in reducing the \chandra\ data, and the referee, Martin
Hardcastle, for helping to improve the paper.  PEJN was partly supported
by NASA grant NAS8-01130.  We acknowledge support from Long Term Space
Astrophysics grant NAG5-11025, Chandra Archival Research grant
AR2-3007X, and contract 81305-001-034V from the Department of Energy
through the Los Alamos National Laboratory.


\end{document}